\documentstyle[psfig,preprint,aps]{revtex}
\tightenlines

\begin{document}
\draft

\title{Average Time Spent by L\'evy Flights and Walks  on
an Interval\\ with Absorbing Boundaries}

\author{
S. V.~Buldyrev$^{1}$,
S.~Havlin$^{1,2}$, A.~Ya.~Kazakov$^{3}$,  M.~G.~E.~da~Luz$^{4}$, 
E.~P.~Raposo$^{5}$, H.~E.~Stanley$^1$, and
G.~M.~Viswanathan$^{1,6}$}

\address{$^{1}$Center for Polymer Studies and Department of Physics,
Boston University, Boston, MA 02215}

\address{$^2$Gonda-Goldschmied Center and Department of Physics, Bar
Ilan University, Ramat Gan, Israel} 

\address{$^3$Mathematical Department, State University of Aerospace 
Instrumentation, 190000, Bol'shaya Morskaya, 67, S.-Petersburg, Russia}

\address{$^4$Departamento de F\'\i sica,
Universidade Federal do Paran\'a, 81531-990, Curitiba--PR,
Brazil}

\address{$^5$Laborat\'orio de F\'{\i}sica Te\'orica e Computacional,
Departamento de F\'{\i}sica, Universidade Federal de
Pernambuco, 50670-901, Recife--PE Brazil} 

\address{$^{6}$Departamento de F\'\i sica,
Universidade Federal de Alagoas,
57072-970, Macei\'o--AL, Brazil}

\date{ EZ7284~~~~ 19 May 2001 }

\maketitle

\begin{abstract}
We consider a L\'evy flyer of order $\alpha$ that starts from a point
$x_0$ on an interval $[O,L]$ with absorbing boundaries.  We find a
closed-form expression for the average number of flights the flyer
takes and the total length of the flights it travels before it is
absorbed.  These two quantities are equivalent to the mean first
passage times for L\'evy flights and L\'evy walks, respectively.
Using fractional differential equations with a Riesz kernel, we find
exact analytical expressions for both quantities in the continuous
limit. We show that numerical solutions for the discrete L\'evy
processes converge to the continuous approximations in all cases
except the case of $\alpha \rightarrow 2$, and the cases of $x_0
\rightarrow 0$ and $x_0 \rightarrow L$. For $\alpha>2$, when the
second moment of the flight length distribution exists, our result is
replaced by known results of classical diffusion.  We show that if
$x_0$ is placed in the vicinity of absorbing boundaries, the average
total length has a minimum at $\alpha=1$, corresponding to the Cauchy
distribution.  We discuss the relevance of this result to the problem
of foraging, which has received recent attention in the statistical
physics literature.
\end{abstract}

\section{Introduction}

\noindent
In the past two decades, L\'evy flights and L\'evy walks (see
\cite{book1,book2,klafter,zolotarev,tsallis,west1,shles,west2,zaslav2,kutner}) 
found numerous
applications in natural sciences.  Realizations of L\'evy flights in
physical phenomena are very diverse, including fluid dynamics,
dynamical systems, and statistical mechanics. 

In general, L\'evy flights and L\'evy walks model anomalous diffusion,
which is governed by rare but extremely large jumps of diffusing
particles. Both L\'evy walks and L\'evy flights are characterized by
broad distributions of their step lengths, for which the second moment
does not exist.  L\'evy walks and L\'evy flights of order
$\alpha<2$ have distributions of step lengths with diverging moments
of order $m \geq \alpha$ and converging moments of order
$m<\alpha$. Hence, the classical central limit theorem, which governs
the behavior of the Brownian motion, is not applicable. According to
the generalized central limit theorem\cite{zolotarev,taqqu}, the
probability density $\Psi(x,n)$ of the displacement $x$ of L\'evy
flights converges after many steps to the L\'evy stable distribution
of order $\alpha$:
\begin{equation}
\label{e00}
\Psi (x,n) = {1 \over \pi}\int_0^\infty
\exp(-n\ell_0^\alpha q^\alpha)\cos(qx)dq, 
\end{equation}
where $\ell_0$ is the characteristic width of the distribution
of a single step and $n$ is number of steps.  This distribution is a
generalization of the Gaussian distribution, and is characterized 
for asymptotically large displacements by
the power law decay of its density with the exponent 
\begin{equation}
\label{e00a}
\mu=\alpha+1.
\end{equation}
There are several definitions of L\'evy walks and L\'evy flights,
which differ in terms of their spatial-temporal correlations (see
e.g. \cite{kutner}). Here we will restrict ourselves to the definition
\cite{shles}. Accordingly, we assume that for L\'evy flights the duration of
each step is constant, so that velocity is proportional to the step
length. Hence, the time of travel is proportional to the number of
steps. Consequently, for L\'evy flights, the mean-square displacement does not
exist as a function of time. This property impedes direct applications
of L\'evy flights to physical phenomena. 

In L\'evy walks, walkers travel with constant velocity, which is
independent of the step length. Hence, the time of travel is
proportional to the total path length. Consequently, the mean-square
displacement exists as function of time, but grows faster than
linearly. This property makes L\'evy walks applicable for modeling
superdiffusion. However, the time evolution of L\'evy flights is
simpler than that of L\'evy walks.  Hence, in the following we will
derive our results for L\'evy flights, keeping in mind that the total
path length of the L\'evy flights corresponds to total time of travel
in L\'evy walks.  In the continuous limit, L\'evy flight process is
described by the superdiffusion equation, which includes integer first
order derivative with respect to time and fractional Riesz operator
with respect to spatial coordinates
\cite{klafter,zolotarev,zaslav,fog,jesp,saichev,gitter}. Here we will
restrict our study to only this class of equations.

Note that usually anomalous diffusion is modeled by the equations of
the Schneider-Wyss type \cite{schneider}, which include fractional
derivatives with respect to time and usual Laplace or Fokker-Planck
operators with respect to spatial coordinates
\cite{klafter,zolotarev,barkai,mbk,met1,met2,rang,barkai1,podlub}. In
this case, the presence of absorbing boundaries can be treated the
same way as in normal diffusion, since after the separation of
variables, the solution can be expressed as series of the usual
eigenfunctions of the boundary problem for the Laplace or
Fokker-Planck operator \cite{gitter,met2,rang,barkai1}.

In the absence of boundaries, the generalized central limit theorem
allows us to treat L\'evy flight diffusion with the help of fractional
differential equations
\cite{klafter,zolotarev,west1,zaslav2,zaslav,fog,jesp,saichev}.
In the presence of boundaries, the validity of fractional derivative
formalism is less clear. Note that the problem of the discrete L\'evy
flights is finite, since it involves the characteristic width $\ell_0$
of the distribution of discrete steps. Thus the problem of L\'evy
flights in the finite domain of linear size $L$ must depend on the
ratio $M \equiv L/\ell_0$. The transition from discrete L\'evy flights to the
fractional differential equation involves transition $\ell_0
\rightarrow 0$.  Consequently, the total number of flights and total
path length diverge as powers of $M$. Since the same is true for
the total path length of the Brownian walker, this problem can be
solved by introducing fractional diffusion coefficient, the same way
it is done in the usual diffusion equation. We will address this point in
Section IV.

L\'evy flights in a slab geometry with absorbing boundaries have been
used to model the transmission of light through cloudy atmosphere \cite{davis}. 
Using heuristic arguments confirmed by
numerical simulations, ref. \cite{davis} found the scaling behavior of the
transmission probability of a photon through a slab of
width $L$ and the total geometrical path length of transmitted and
reflected light. This behavior was experimentally observed in
\cite{pfe}. We analytically derive an exact expression for the
transmission probability in Section IV.

Very recently \cite{gitter}, the approximate expressions for the mean first passage
time for both Schneider-Wyss sub-diffusion equation  
\cite{schneider} and superdiffusion equation\cite{fog} have been
obtained  by separation of variables.  The latter case
exactly corresponds to the L\'evy flight problem, which we treat here
(see Section IV).  The problem of L\'evy walks on a finite interval with
absorbing boundaries has already been addressed in \cite{rob}.  In
that paper, the authors used an integral equation approach and
performed the Laplace transform in the temporal domain. They found the
asymptotic behavior of the survival probability, which is related to
the asymptotic behavior of the first passage time.  An alternative
approach to the L\'evy walk problem which employs the fractional Kramers
equations can be found in \cite{barkai}. However, as far as we know,
exact expressions for the mean first passage time for L\'evy
flights and L\'evy walks have not been derived.

The structure of the paper is the following.
In Sections II and III, we find the mean time of travel before
absorption for both discrete L\'evy flights and L\'evy walks as
solutions of Fredholm integral equations of the second kind
\cite{smirnov} with a power-law kernel truncated by a cut-off at small
distance $\ell_0$. 

In Section IV, we treat these equations in the limit $\ell_0
\rightarrow 0$ and reduce our problem to the solution of fractional
differential equations with Riesz kernels
\cite{klafter,zaslav,podlub,minsk}. These equations were previously
applied in the plane contact problem \cite{arutun} of linear creep
theory and were solved using spectral relationships with Jacoby
polynomials \cite{podlub,popov} and by the Sonin inversion formula
\cite{popov,sonin}. We also show how our method is related to the
method of separation of variables first applied to the partial
fractional differential equation of superdiffusion in \cite{gitter}.

In Section V, we compare our analytical solutions
obtained in the continuous limit $\ell_0 \rightarrow 0$ with the
numerical solutions of the Fredholm equations obtained for discrete
L\'evy flights.  We show that fractional differential
equations can serve as good approximations for L\'evy flights
with absorbing boundaries for $\alpha<2$. We also show
that these approximations break down when $\alpha\rightarrow 2$ and in
the vicinity of the absorbing boundaries.

Finally, in Section VI, we discuss the relevance of our results to the
problem of biological foraging.  Recently, L\'evy flights have been
used to model animal foraging \cite{shles,levand,cole,vis1,vis2,vis3} and cell diffusion
\cite{upad}.  It
has been suggested \cite{vis2} that L\'evy flights with $\alpha=1$
provide the optimal strategy of foraging in case of sparse food sites,
if any food site can be revisited.  This suggestion was
based on the optimization of foraging efficiency, defined to be
the inverse of the average total path length of the flyer before
the flyer is absorbed by traps randomly distributed with certain
density in $d$-dimensional space.  The average total path length has been
approximated \cite{vis2} as a product of the average length of a
single flight and the average number of flights before the flyer is
absorbed by traps.  It has been shown \cite{vis2} that this product
has a maximum at $\alpha=1$, if the starting point of the flyer $x_0$
is selected in the vicinity of the absorbing boundary.  Here we
confirm this result in the one-dimensional case using both an analytical
expression for the average total length of flights obtained in the
continuous limit and the numerical solution for the discrete L\'evy
process.  We show that for the case of $x_0$ in the vicinity
of the absorbing boundary, discrete and continuous
solutions have the same power law asymptotic behavior, but
their amplitudes are different. As a consequence, the continuous limit 
approximation has an additional minimum at $\alpha \rightarrow 2$,
which is absent in the discrete case. 
This finding indicates that the fractional differential equation approach
to L\'evy flights breaks down in the vicinity of the absorbing boundary.

In Appendix A, we derive the fractional differential operator for the
L\'evy flight problem with absorbing boundaries. In Appendix B, we
derive exact analytical expressions for the number of flights and the
total length traveled before absorption, using the Sonin inversion formula
for the Riesz fractional equation.

\section{Mean Number of Flights}

Consider a L\'evy flight that starts at point $x_0$ of the interval
$[0,L]$ with absorbing boundaries. The flyer makes independent
subsequent flights of variable random lengths $\ell$ with equal
probability in both directions.  The length of each flight is taken from
the power law distribution 
\begin{equation}
\label{e02}
P(|\ell|>r)=(\ell_0/r)^\alpha,
\end{equation} 
where the
exponent $\alpha$ can vary between 0 and 2, and $\ell_0$ is the minimal
flight length, which serves as lower cutoff of the distribution. The
probability density of the flight length is given by
\begin{equation}
\label{e0}
p(\ell)={\alpha\ell_0^\alpha\over
2}{\theta(|\ell|-\ell_0)\over
|\ell|^{\alpha+1}},
\end{equation}
where $\alpha\ell_0^\alpha/2$ is a normalization constant and
$\theta(x)=1$ for $x>0$ or $0$ otherwise.  The exponent
$\mu$ of refs. \cite{vis1,vis2} is identical to $\alpha+1$. When
$\alpha>2,$ the second moment of the flight distribution converges and
the process becomes equivalent to normal diffusion. As soon as the
flyer lands outside the interval $[0,L]$, the process is
terminated. Instead of the probability density Eq. (\ref{e0}) one
could use any power-law decaying density \cite{zolotarev,rob} regularized at
small distances $\ell_0$, including L\'evy stable distribution Eq.
(\ref{e00}) with $n=\pi/[2\Gamma(\alpha)\sin(\pi\alpha/2)]$, where
$\Gamma(\alpha)\equiv\int_0^\infty t^{\alpha-1}e^{-t}dt$
is Euler 
$\Gamma$-function \cite{stegun}. This value of $n$ is selected so that the asymptotic behavior of Eq.
(\ref{e00}) coincides with Eq. (\ref{e0}) \cite{zolotarev}.
We use a truncated power-law density for the
simplicity of analytical treatment.

We are interested in two quantities: the average number of flights
before absorption $\langle n \rangle$ and the average total distance
$\langle S\rangle$ traveled before absorption. Note that we consider the
length of the last flight to be equal to the distance from the previous
landing point to the boundary of the interval which flyer crosses during
its last flight. This condition makes our problem equivalent to the
problem of L\'evy walks \cite{rob} with the time defined to be
equal to the sum of the flight lengths.

Suppose that the probability density of finding a L\'evy flyer at point
$x$ after $n$ flights is $P_n(x)$. Then the probability density after
$n+1$ flights is given by the convolution of the probability density
$P_n(x)$ and the probability density of the next flight $p(x)$ given by
Eq.~(\ref{e0})
\begin{equation}
\label{e1}
P_{n+1}(y)=\int_0^L {p(y-x)P_n(x)dx}.
\end{equation}

Let ${\cal L}_\alpha(\ell_0)$ be an integral operator with kernel
$p(x-y)$ which is defined on a function $f(x)$ of an interval $[O,L]$
as
\begin{equation}
\label{e2}
[{\cal L}_\alpha f](y)\equiv{\alpha\ell_0^\alpha\over
2}\int_0^L{f(x)\theta(|y-x|-\ell_0)dx\over
|y-x|^{\alpha+1}}.
\end{equation}
One can see, that ${\cal L}_\alpha$ is a self-conjugate operator with
respect to a scalar product $(f,g)=\int_0^L f(x)g(x)dx$.  It can be
shown that for any continuous function $f$, $\int_0^L|{\cal L}_\alpha
f|dx \leq [1-(2/M)^{\alpha}]\int_0^L|f|dx$, where $M=L/\ell_0$. The
value $[1-(2/M)^\alpha]$ can be regarded as the norm of the operator
${\cal L}_\alpha$.  It has a physical meaning of the survival
probability, i.e. the probability for the flight that starts at the
center of the interval to stay unabsorbed, which is less than one.

The
distribution after $n$ flights is given by
\begin{equation}
\label{e3}
P_n(x)=[{\cal L}_\alpha^nP_0](x).
\end{equation}
The initial probability density of the flyer located at
position $x_0$ is the Dirac delta function, $P_0(x)=\delta(x-x_0)$.  The
probability that the flyer remains unabsorbed after $n$ flights, 
\begin{equation}
\label{e4}
\int_0^L[{\cal L}_\alpha^n P_0](x)dx<\left[1-(2/M)^{\alpha}\right]^n
\end{equation}
decays exponentially with $n$.
The probability that the flyer is absorbed exactly on the
$n^{\mbox{\scriptsize th}}$ flight is
\begin{equation}
\label{e5}
P_n=\int_0^L\left[\left({\cal L}_\alpha^{n-1}-{\cal
L}_\alpha^n\right)P_0\right](x)dx, 
\end{equation}
and thus, the average number of flights spent by the flyer on the
interval is
\begin{equation}
\label{e6a}
\langle n\rangle=\sum_{n=1}^\infty P_nn=\int_0^L \sum_{n=0}^\infty[{\cal
L}_\alpha^n P_0](x)dx=-\int_0^L\left[({\cal L}_\alpha-{\cal I})^{-1}P_0\right]
(x)dx.
\end{equation}
The infinite sum in Eq. (\ref{e6a}) converges, since the norm of ${\cal
L}_\alpha$ is less than one. Here $({\cal L}_\alpha-{\cal
I})^{-1}$ is the inverse operator with respect to the operator ${\cal
L}_\alpha-{\cal I}$, and ${\cal I}$ is the unity operator. Since the
operator $({\cal L}_{\alpha}-{\cal I})^{-1}$ is also a self-conjugate
operator, and $P_0(x)=\delta(x-x_0)$, we have
\begin{equation}
\label{e6}
\langle n\rangle=
\left[({\cal L}_\alpha-{\cal I})^{-1}h\right](x_0),
\end{equation}
where $h(x)=-1$ is the constant function.
This equation can be rewritten as a Fredholm integral equation of the
second kind \cite{smirnov} with the kernel $p(x_0-x_1)$:
\begin{equation}
\label{e2.4}
\langle n(x_0)\rangle=1+\int_0^L\langle n(x_1)\rangle p(x_0-x_1)dx_1. 
\end{equation}

Equation (\ref{e2.4}) can be interpreted as a recursive method to
determine $\langle n(x_0)\rangle$.  Indeed, the average total number
of flights for the process that start at $x_0$ is equal to one
(contribution from the very first flight, which always takes place)
plus the convolution of the average total number of flights for the
processes that start from all possible landing points $x_{1}$ of the
first flight inside the interval and the probability density
$p(x_0-x_1)$ to land at these points after the first flight.
 
In general, consider a quantity $Q(x_0)=\langle \sum_{i=1}^\infty q_i \rangle$,
where $q_i=q(x_{i-1},x_i)$ is a function of the starting point $x_{i-1}$ and ending point $x_i$ of the 
$i$-th flight, and $\langle \rangle$ denotes the average over all possible processes
starting at point $x_0$. Then, in analogy with the average number of flights, 
such a quantity must satisfy a recursion relation
\begin{equation}
\label{e2.4a}
Q(x_0)=\langle q_0(x_0)\rangle +\int_0^L Q(x_1) p(x_0-x_1)dx_1, 
\end{equation}
where $\langle q_0(x)\rangle\equiv
\int_{-\infty}^{\infty}p(x_1-x_0)q(x_0,x_1)dx_1$. Note that if $x_1$
is outside the interval $[0,L]$, the particle is absorbed by one of
the boundaries and the value $q(x_0,x_1)$ should be defined according to
its physical meaning for the absorbed particle. Equation (\ref{e2.4a})
is identical to Eq.(\ref{e6}) with $h(x)=-\langle q_0(x) \rangle$.

As an example of application of Eq.~(\ref{e2.4a}), let us consider
quantity $Q(x_0)$ to be the total flux through the right boundary.
This flux is related to the transmission probability of photons
through the clouds\cite{davis}.  By definition, the flux through the
right boundary is equal to the probability $P_r(x_0)$ of the
absorption of the flyer that starts at point $x_0$ by the absorbing
boundary $x=L$.  In this case, quantity $q$ must be defined as
$q(x_0,x_1)=\theta(x_1-L)$.  The very first flight is absorbed by the
right boundary with probability $p_r(x_0) \equiv \langle
q_0(x_0)\rangle= \int_L^\infty p(x_1-x_0)dx_1$, or after integration
\begin{equation}
\label{es3}
p_r(x_0)=\left\{ 
\begin{array} {c}
\left[\ell_0/(L-x_0)\right]^\alpha/2, \\
1/2,
\end{array} 
\begin{array} {c}
0 \leq x_0<L-\ell_0 \\
L-\ell_0 \leq x_0 \leq L
\end{array}
\right . .
\end{equation} 
Therefore, $P_r(x_0)$ satisfies Eq.~(\ref{e2.4a}) with 
$\langle q_0(x_0)\rangle=p_r(x_0)$ and Eq.~(\ref{e6}) with $h(x_0)=-p_r(x_0)$.
In the next section, we will apply this method to define the total
path length of the flyer.

\section{Average total path length traveled by the flyer}

The average total path length traveled by a L\'evy flyer before
absorption is equivalent to the total time spent by a L\'evy walker
before absorption \cite{rob}. The evolution of the probability
density of L\'evy walks was studied \cite{rob} 
in terms of
time. This approach leads to integral equations involving integration over
time and space. Here we restrict our study to the problem of the
average total path length before absorption of discrete L\'evy
flights. This particular problem can be solved in much simpler
terms. 

In the absence of the absorbing boundaries, the average flight
length with probability density $p(\ell)$ of Eq.~(\ref{e0}) is given
by $\langle | \ell | \rangle =\int_{-\infty}^{+\infty} |\ell| p(\ell)d
\ell$, which is independent of the starting point and diverges for
$\alpha \leq 1$.  In the presence of the absorbing boundaries,
the flight starting from a point $y$ cannot exceed the
distances $y$ and $L-y$ from this point to the boundaries.  One can
show that the average length of a flight that starts from a point $y$
of an interval $[\ell_0,L-\ell_0]$ is given by
\begin{equation}
\label{e2.1}
s(y)\equiv \langle |\ell(y)|\rangle 
={\alpha\ell_0^\alpha\over
2}\left[\int_0^{y-\ell_0}{dx\over(y-x)^{\alpha}}+
\int_{y+\ell_0}^L{dx\over(x-y)^{\alpha}} \right . 
\end{equation}
\[
\left.
+y\int_{-\infty}^0{dx\over(y-x)^{\alpha+1}} + 
(L-y)\int_L^\infty{dx\over(x-y)^{\alpha+1}}\right],
\]
which converges for $\alpha>0$.
If $\alpha \neq 1$, Eq.~(\ref{e2.1}) reduces to
\begin{equation}
\label{e2.2}
s(y)={\ell_0\over 2 (1-\alpha)}
\left[ \left({\ell_0\over y}\right)^{\alpha-1}+
\left({\ell_0\over L-y}\right)^{\alpha-1}-2\alpha\right].
\end{equation}
In case $\alpha = 1$, we have from Eq.~(\ref{e2.1})
\begin{equation}
\label{e2.2a}
s(y) ={\ell_0\over 2}\left[ 
\ln \left({y \over \ell_0}\right)+
\ln \left({L-y \over \ell_0}\right)+
2\right].
\end{equation}
If $0\leq y<\ell_0$, Eq.~(\ref{e2.2}) must be replaced by
\begin{equation}
\label{e2.2b}
s(y) ={y\over 2}+{\ell_0\over 2 (1-\alpha)}
\left[ \left({\ell_0\over L-y}\right)^{\alpha-1}-\alpha\right],
\end{equation}
or by
\begin{equation} 
\label{e2.2c}
s(y) ={L-y\over 2}+{\ell_0\over 2 (1-\alpha)}\left[ 
\left({\ell_0\over y}\right)^{\alpha-1}-\alpha\right],
\end{equation}
if $L-\ell_0 <y \leq L$. Analogous changes must be made in
Eq.~(\ref{e2.2a}).

Thus, according to Eq. (\ref{e2.4a}), 
the total average path length for the process that starts at point $x_0$ is 
\begin{equation}
\label{e2.5}
\langle S(x_0)\rangle=s(x_0)+\int_0^L\langle S(x_1)\rangle p(x_0-x_1)dx_1,
\end{equation}
or
\begin{equation}
\label{e2.3}
\langle S (x_0)\rangle=-\left[({\cal L}_{\alpha}-{\cal I})^{-1}s\right] (x_0).
\end{equation} 
This equation is identical to the Eq.~(\ref{e6}) in which function
$h(x)=-s(x)$. 
We will solve Eqs. (\ref{e2.4}) and (\ref{e2.5}) numerically in Section V.

\section {The Continuous Limit}

Appendix A shows that for $\alpha <2$, operator ${\cal L}_\alpha(\ell_0)-{\cal
I}$ tends to zero when $\ell_0\to 0$ for any function $f(x)$ that has second
derivative $f''(x)$ for $0<x<L$ and
has finite limits $f(0)$ and $f(L)$. It
can also be shown that for such functions and $\alpha<2$ there exists
an operator \cite{klafter,podlub,D}
\begin{equation}
\label{e7}
{\cal D}_\alpha \equiv \lim_{\ell_0\to 0}\ell_0^{-\alpha}[{\cal
L}_\alpha(\ell_0)-{\cal I}]. 
\end{equation}
The result of this operator acting on any such function is defined as
\begin{equation}
\label{e7c}
[{\cal D}_\alpha f](y)=\mbox{V.P.}\int_0^L{\mbox{sgn}(x-y)f'(x)dx\over
2|y-x|^\alpha} - {f(0) \over 2y^\alpha} - {f(L) \over 2(L-y)^\alpha},
\end{equation}
where $f'(x)$ is the first derivative of the function $f(x)$.  This
operator is a self-conjugate operator similar to the double
differentiation operator $d^2/dx^2$. It can be expressed \cite{klafter,podlub}
as the linear combination of right and left Riemann-Liouville fractional
derivatives of the order $\alpha$.  The difference of the two operators
\begin{equation}
\label{e7a}
d_\alpha(\ell_0) \equiv \ell_0^{-\alpha}[{\cal
L}_\alpha(\ell_0)-{\cal I}]-{\cal D}_\alpha
\end{equation}
decays as $\ell_0^{2-\alpha}$ when $\ell_0 \rightarrow 0$. 
Appendix A also shows that the leading term of the operator $d_\alpha$ is
proportional to the operator of the second derivative
\begin{equation}
\label{e7b}
d_\alpha(\ell_0) = \ell_0^{2-\alpha}{\alpha\over 2(\alpha-2)}{d^2\over dx^2}
+o(\ell_o^{2-\alpha}).
\end{equation}

In analogy with the diffusion equation with continuous time, we can
define a superdiffusion equation \cite{klafter,zaslav,fog,gitter}
based on L\'evy flights. Instead of the discrete process defined by
Eq.~(\ref{e1}), one can write
\begin{equation}
\label{e9}
{\partial P(x,t)\over\partial t}={\ell_0^\alpha\over t_0}{\cal D}_\alpha
P(x,t),
\end{equation}
where $t_0$ is the duration of each flight, and $\ell_0^\alpha/t_0$ is
the fractional analog of the diffusion coefficient. Note that $\ell_0$
plays a role similar to the mean free path, and $t_0$ plays the role
of the mean collision interval. 

The operator ${\cal D}_\alpha$ has an orthogonal normalized set of
eigenfunctions $f_k(x)$, such that ${\cal D}_\alpha f_k(x)=\lambda_k
f_k(x)$, and $f_k(0)=f_k(L)=0$ \cite{minsk,kogan}. 
Similarly to the solution of usual diffusion equation, the solution of
Eq.~(\ref{e9}) can be expressed via separation of variables as a series
of eigenfunctions
\begin{equation}
\label{e10}
P(x,t)=\sum_{k=1}^\infty e^{\lambda_k\ell_0^\alpha t/t_0}f_k(x)
\int_0^Lf_k(y)P(y,0)dy.
\end{equation} 
In Ref. \cite{gitter}, where the method of separation of variables for
the superdiffusion equation on a finite interval has been first
proposed, it has been assumed that the eigenvalues $\lambda_k$
asymptotically behave at large $k$ as $\lambda_k\sim-(k/L)^\alpha<0$
and that the eigenfunctions $f_k(x)$ can be well approximated by the
eigenfunctions $\sqrt{2/L}\sin(x\pi k/L)$ of the Laplace operator with
absorbing boundary conditions. Numerical studies \cite{bg} confirm these
assumptions but show that eigenfunctions $f_k$ and sines have different 
behavior near absorbing boundaries, namely, $f_k(x) \sim x^{\alpha/2}$ as
$x\rightarrow 0$. 

Having defined the properties of the operator ${\cal D}_\alpha$, we can
derive the closed form expression for the average time spent by the
continuous L\'evy flight process on the interval. Formal substitution of
Eq.~(\ref{e7}) into Eq.~(\ref{e6}) yields
\begin{equation}
\label{e11}
\langle t\rangle=t_0\langle
n\rangle
={t_0\over\ell_0^\alpha}\left[{\cal D}_\alpha^{-1}h\right](x_0)
={t_0\over\ell_0^\alpha}g(x_0),
\end{equation}
where function $g(x)$ satisfies the equation
\begin{equation}
\label{e14}
{\cal D}_\alpha g(x)=h(x)=-1.
\end{equation}
Note that $g(x)$ has to satisfy boundary conditions $g(0)=0$, $g(L)=0$.
Otherwise, according to Eq.~({\ref{e7c}), the right hand side of 
Eq.~(\ref{e14}) would contain singularities. In the general case, the
equation 
\begin{equation}
\label{e14a}
\mbox{V.P.}\int_0^L{\mbox{sgn}(y-x)f'(y)dy\over
2|y-x|^\alpha} = h(x)
\end{equation}
with absorbing boundary conditions
\begin{equation}
\label{e14b}
f(0)=f(L)=0
\end{equation}
belongs to a known class of generalized Abel integral equations with
Riesz fractional kernel \cite{podlub,minsk,arutun,popov}. It can be
shown \cite{minsk,popov} that such an equation with boundary
conditions (\ref{e14b}) has a unique solution which can be obtained
via spectral relationships for Jacobi polynomials \cite{podlub,popov}
or by the Sonin inversion formula \cite{popov,sonin} (see Appendix B).
Similar inversion formulae are given in ref. \cite{minsk}. 
In the case $h(x)=-1$, the solution can be expressed in elementary
functions:
\begin{equation}
\label{e12}
g(x)={2\sin(\pi\alpha/2)\over\pi\alpha}[(L-x)x]^{\alpha/2}.
\end{equation}
One can verify this solution by performing contour integration around
the cut $[0,L]$ on the complex plane and computing the residue of
the integrand at infinity.  
It should be pointed out that $g(x)$ can be expanded in a series of
eigenfunctions $f_k(x)$:
\[
g(x_0)= -\sum_{k=1}^\infty \lambda_k^{-1}f_k(x_0)\int_0^L f_k(x)dx.
\] This expansion is similar in spirit to an approximation,
obtained in \cite{gitter}, where the exact eigenfunctions were approximated
by sines. Although approximation \cite{gitter} correctly predicts the
power law dependence $g(x_0)\sim L^\alpha$ for the points $x_0$ in the
center of the interval, it differs from Eq. (\ref{e12}) in the proportionality
coefficient and in the behavior near
absorbing boundaries \cite{bg}.

Note that for $\alpha >2$, Eq.~(\ref{e9}) should be replaced \cite{D}
by the standard diffusion equation with diffusion coefficient
$D={\ell_o^2 \alpha/[t_o 2(\alpha -2)]}$. In this case, the average
time spent by the flyer before absorption is given by the classical
equation $\langle t \rangle=x(L-x)/(2D)$. Note that
the diffusion coefficient $D$ resembles the proportionality
coefficient in Eq.~(\ref{e7b}).

One can argue that the expression (\ref{e12})  may yield the average 
number of flights taken by the discrete L\'evy flight
process in the limit of $\ell_0\rightarrow 0$.  Indeed, according to
Eq.~(\ref{e7a})
\begin{equation}
\label{e16}
{\cal L}_\alpha(\ell_0)-{\cal I}=\ell_0^{\alpha}[{\cal D}_\alpha+
d_\alpha(\ell_0)],
\end{equation}
where operator $d_\alpha(\ell_0) \rightarrow 0$, as $\ell_0\rightarrow
0$. Formally expanding $({\cal L}_\alpha -{\cal I})^{-1}$ in powers of
$d_\alpha$, we obtain
\begin{equation}
\label{e15a}
[{\cal L}_\alpha-{\cal I}]^{-1}=\ell_0^{-\alpha}({\cal D}_\alpha^{-1}
-{\cal D}_\alpha^{-1}d_\alpha{\cal D}_\alpha^{-1} +...),
\end{equation}
and thus
\begin{equation}
\label{e15}
\langle n \rangle=
\ell_0^{-\alpha}\left\{ g(x_0)-[{\cal D}_\alpha^{-1}d_\alpha g](x_0)
+...\right\}
\end{equation}
Note that expansion (\ref{e15}) is formal and may not converge. 
We will test this assumption numerically in Section V.
In order to distinguish 
the average number of flights for the discrete process, $\langle n \rangle$,
from the continuous limit approximation, we will denote the latter by 
$n_\alpha(x)$:
\begin{equation}
\label{e15b}
n_\alpha(x_0)\equiv \ell^{-\alpha}g(x_0)=
{\sin(\pi\alpha/2)\over\pi\alpha/2}
\left[(L-x_0)x_0 \over \ell_0^2\right]^{\alpha/2}=
{\sin(\pi\alpha/2)M^\alpha(z-z^2)^{\alpha/2}\over\pi\alpha/2},
\end{equation}  
where $z \equiv x_0/L$, $M \equiv L/\ell_0$.

Analogously, we will denote the continuous limit approximation for
the average total path length by $S_\alpha$.
In the continuous limit, Eq.~(\ref{e2.3}) should be replaced by
Eq.~(\ref{e14a}) with $f(y)=S_\alpha(y)$, $h(x)=\ell_0^{-\alpha}s(x)$, and
absorbing boundary conditions (\ref{e14b}).  In this case (See
Appendix B), Sonin formula leads to an expression containing
hypergeometric functions:
\begin{equation}
\label{e9.b}
S_\alpha(x_0)={L(2-\alpha)\over 2(1-\alpha)}\left[1-4{\psi_\alpha(z)+\psi_\alpha(1-z)
\over \alpha(\alpha+2)\mbox{B}({\alpha \over 2},{\alpha \over 2})}\right]+
{2LM^{\alpha-1}\sin\left({\pi\alpha\over 2}\right)
(z-z^2)^{\alpha\over 2}\over \pi (\alpha -1)},
\end{equation}
where $\mbox{B}(a,b)\equiv\Gamma(a)\Gamma(b)/\Gamma(a+b)$ is Euler
$\mbox{B}$-function, 
\[
\psi_\alpha(z)=
F(2-{\alpha\over 2},{\alpha\over 2},{\alpha\over 2}+2,z)z^{{\alpha\over 2}+1},
\]
and $F$ is the hypergeometric function \cite{stegun}
\[
F(a,b,c,x)\equiv{\Gamma(c)\over\Gamma(a)\Gamma(b)}\sum_{n=0}^\infty
{ \Gamma(n+a)\Gamma(n+b)x^n\over\Gamma(n+1)\Gamma(n+c)}. 
\]
In case $\alpha=1$, corresponding to the Cauchy distribution, the hypergeometric
function $\psi_\alpha(z)$ can be expressed in elementary functions
$\psi_1(z)={3\over 4}\left[{\pi\over 2}+\sin^{-1}(2z-1)
-2\sqrt{z-z^2}\right]$
and Eq.~(\ref{e9.b}) yields $S_1(x_0)=L{2\over\pi}\sqrt{z-z^2}\ln M +O(1)$,
where terms $O(1)$ do not depend on $M$ and can be found from Eq. (\ref{e9.b})
by L'H\^opital's rule.

Note that the average total path length traveled before absorption by
the L\'evy flyer can be expressed in terms of survival probability
$\Theta(t)$ of the L\'evy walker: $\langle S\rangle=
\int_0^\infty\Theta(t)dt$. According to \cite{rob}, this survival
probability exhibit for $t\rightarrow \infty$ asymptotic exponential
decay $\Theta(t)\sim \exp (-|\Lambda_1|t)$, where $\Lambda_1$ is the main
eigenvalue of the correspondent problem. Substituting $\Theta(t)$ by
its asymptotic, we can estimate $ \langle S\rangle \sim {1 /
|\Lambda_1|}$.  Thus $ \langle S\rangle$ and ${1 / |\Lambda_1|}$
must have identical asymptotic behavior for large $L$. Indeed,
Eq.~(\ref{e9.b}) yields $S_\alpha(x) \sim L$ for $0<\alpha<1$ and
$S_\alpha(x) \sim L^\alpha$ for $1\leq\alpha<2$, in complete agreement
with asymptotic approximations of \cite{rob}.

Finally, we will find the probability $P_r(x_0)$ of the absorption by
the right boundary in the continuous limit. According to Eqs.~(\ref{e2.4a})
and (\ref{es3}),
$P_r(x_0)$ should, in continuous case, satisfy Eq. (\ref{e14}) with
$h(x_0)=-\ell_0^{-\alpha}p_r(x_0)$, i.e.
\begin{equation}
\label{e17.b}
\cal{D}_\alpha P_r(x_0)=-(L-x_0)^{-\alpha}/2.
\end{equation}
For $x_0=0$, the flyer is immediately absorbed by the left boundary,
so $P_r(0)=0$. For $x_0=L$, the flyer is immediately absorbed by the
right boundary, so $P_r(L)=1$. Thus the second term in the expression
(\ref{e7c}) for $\cal{D}_\alpha P_r(x_0)$ is equal to zero and the
third term cancels out with the right hand side. Hence $P_r(x_0)$
satisfies homogeneous Eq.~(\ref{e14a}) and boundary conditions
$P_r(0)=0$, $P_r(L)=1$. This solution can be expressed in terms of the 
homogeneous solution $\varphi_0(x)$ obtained in Appendix B:
\begin{equation}
\label{e16.b}
P_r(x_0)={\int_0^{x_0}\varphi_0(y)dy\over\int_0^L\varphi_0(y)dy}
=\left({x_0\over L}\right)^{\alpha\over 2}{F({\alpha\over 2},1-{\alpha\over 2},{\alpha\over 2}+1,
{x_0\over L}) \over {\alpha \over 2} B({\alpha\over 2}, {\alpha\over 2})},
\end{equation}
Note that the probability of the absorption by the left boundary,
$P_l(x_0)=P_r(L-x_0)= 1-P_r(x_0)$.  For $x_0\rightarrow 0$, the
asymptotic behavior of $P_r(x_0)$ is given by 
$P_r(x_0) \sim (x_0 / L)^{\alpha / 2}$ which is in complete agreement with the
result of ref. \cite{davis} for the transmission probability of the
photons through the clouds of depth $L$.

\section {Numerical solution}

The goal of this section is to treat Eq.~(\ref{e2.4}) and
Eq.~(\ref{e2.5}) numerically and to compare the results with the
continuous limit solutions Eq.~(\ref{e15b}) and Eq.~(\ref{e9.b}).  To
perform numerical integration of Eqs.(\ref{e2.4}) and (\ref{e2.5}), we
replace the integration by summation and the kernel $p(x-y)$ by the matrix
$A_{ij}$, $0<i<M$, such that $A_{ii}=0$ and 
\begin{equation}
\label{e3.1}
A_{ij}={1 \over 2}\left[ {1\over |i-j|^{\alpha}}- {1\over
(|i-j|+1)^{\alpha}}\right],\qquad i\neq j.
\end{equation}
Accordingly, the average flight length
$s(x)$ performed from the point $k=x/\ell_0$ is
replaced by $M-1$-dimensional vector $\vec{s}$ with elements
\begin{equation}
\label{e3.2}
s_k={\ell_0 \over 2(1-\alpha)}\left[ 
{1\over k^{\alpha-1}}+ {1\over (M-k)^{\alpha-1}} -2\alpha\right].  
\end{equation}
The average number of flights for the process that starts from point
$k=x/\ell_0$ is
\begin{equation}
\label{e3.3}
{\langle n \rangle}_k=\left(\sum_{m=0}^\infty A^m\vec{ e_k} \cdot
\vec{c}\right)=\left[(I-A)^{-1}\vec{c}\right]_k,
\end{equation}
where $\vec{c}$ is the vector with all components equal to 1, and
$\vec{e_k}$ is a unit basis vector with components $e_{ki}=\delta_{ki},$
where $\delta_{ki}$ is the Kronecker delta.  Analogously, average total
length is equal to
\begin{equation}
\label{e3.4}
\langle S \rangle_k=\ell_0\left(\sum_{m=0}^\infty A^m\vec{ e_k} \cdot 
\vec{s}\right)=\left[(I-A)^{-1}\vec{s}\right]_k.
\end{equation}
The symmetric matrix $R=(A-I)^{-1}$ is the analog of the self-conjugate
operator $({\cal L}-{\cal I})^{-1}$.
Using iterative techniques for matrix inversion, we obtain the numerical
solutions for $\langle n \rangle$ and $\langle S \rangle$. 

In Figure 1, we compare the numerical solution 
(\ref{e3.3}) for $\langle n(x_0) \rangle$
and the continuous limit approximation $n_\alpha(x_0)$ given by Eq.~(\ref{e15b})
for the case $x_0=L/2$.
In order to test the asymptotic convergence, we have to divide both
functions by $(M/2)^\alpha$.  
It can be seen that for $x_0/\ell_0\gg 1$, Eq. (\ref{e15b}) 
provides good approximation for the average number of flights
of the discrete process defined by Eq.~(\ref{e6}).  Studying the
difference between the numerical values of $\langle n\rangle
(M/2)^{-\alpha}$ and the continuous approximation
$2\sin(\pi\alpha/2)/(\pi\alpha)$ for $x_0=L/2$, we confirm that for $2>
\alpha >1$ this difference decays as $M^{\alpha -2}$ with $M\rightarrow
\infty$. This is in
agreement with Eq.~(\ref{e7b}). However, for $\alpha <1$ the difference
between the numerical solution and the continuous approximation converges as
$M^{-1} > M^{\alpha -2}$. The term $M^{-1}$ is proportional to the error
of replacement of integration in Eq.~(\ref{e2.4}) by summation. 

In Fig.~2, we compare the numerical solution (\ref{e3.4}) for $\langle
S(x_0) \rangle$ and the continuous limit approximation $S_\alpha(x_0)$ given
by Eq.~(\ref{e9.b} ) for
the case $x_0=L/2$.  In order to test the asymptotic convergence in
this case, we have to divide both functions by
$(LM^{\alpha-1}-L)/(\alpha-1)$.  It can be seen that for
$x_0/\ell_0\gg 1$, Eq.~(\ref{e9.b}) provides good
approximation for the average total traveled  length in the discrete
process defined by Eq.~(\ref{e2.3})

Now we will examine the quality of the continuous limit approximation
in the vicinity of the absorbing boundary. For simplicity we will study
only the behavior of the average number of flights. The approximation
for the total path length has similar problems.
Figure~3 shows that for $x_0/\ell_0 = 1$, the correction terms in
Eq.~(\ref{e15}) cannot be neglected. As shown in Appendix A, the
operator $d_\alpha\to 0$ for any fixed $x_0$ if $\ell_0\rightarrow 0$,
but does not vanish if $x_0$ and $\ell_0$ both approach zero, so that
their ratio $x/\ell_0 \rightarrow r>0$.  Accordingly, the value of
$\langle n\rangle$ behaves for $x_0/\ell_0=r$, $M \rightarrow \infty$
as $\chi(\alpha)(rM)^{\alpha/2}$, where $\chi(\alpha) >
2\sin(\pi\alpha/2)/(\pi\alpha)$ is some unknown function that can be
estimated numerically (see Fig.~3).  It is likely that $\chi(\alpha)$
remains positive as $\alpha \rightarrow 2$.  The analytical determination
of the function $\chi(\alpha)$ remains an unsolved
problem. Nevertheless, continuous approximation correctly predicts the
leading factor $M^{\alpha/2}$ for the average number of flights
started in the vicinity of the absorbing boundary.

In summary, comparison of the numerical solutions for the mean first
passage time of L\'evy flights and L\'evy walks and the exact
solutions of these problems in the continuous case suggests that
fractional differential equation for superdiffusion Eq.~(\ref{e9})
with absorbing boundary conditions provides good approximation for
discrete L\'evy flights on a finite interval with absorbing
boundaries. However this approximation breaks down when $\alpha
\rightarrow 2$ and in the vicinity of the absorbing boundaries.

\section {Analysis of the Total Path Length: Implications for Biological Foraging}  

Recently, biological foraging has been modeled by L\'evy
flights\cite{shles,levand,cole,vis1,vis2,vis3}.  The case of
non-destructive foraging (defined in \cite{vis2} as case in which
``target sites'' can be revisited not just once but many times)
corresponds to $x_0=\ell_0$, i.e.  the forager starts its next search
from the previously-visited food site, located at the origin. The prey
may reappear at this site.  Accordingly, coming back to the origin may
be profitable in terms of foraging efficiency, which is
defined\cite{vis2} as the inverse average total path length before
finding next food site. With help of Monte-Carlo simulations, it has
been shown \cite{vis2} that, in case of non-destructive foraging, the
foraging efficiency has maximum at $\alpha=1$.

We confirm this result, using numerical solution (\ref{e3.4}). Figure
4 shows a semi-logarithmic plot of $\langle S\rangle$ versus $\alpha$
for $x_0=\ell_0$ and various values of $M$. On can see that the
minima, $\alpha_{\min}(M)$, shift towards $\alpha=1$ as $M \rightarrow
\infty$.  Heuristic approximations \cite{vis2,vis3} suggest that foraging
efficiency has a maximum at $1-\epsilon$, where $\epsilon \sim (\ln
M)^{-2}$. Consequently, the average total path length should have
minimum at the same point. Figure 5 confirms this prediction for the
numerical solution. It shows the graph of $\alpha_{\min}(M)$ versus
$[\ln (M)]^{-2}$, which is almost a straight line with an intercept
$\alpha_{\min}(\infty)\approx 1$.

In the following,
we will prove this result using the continuous limit approximation
(\ref{e9.b}).  Accordingly, we will find the behavior of $S_\alpha(x_0)$
for the case when the starting point $x_0$ is selected in the vicinity
of the absorbing boundary and show that in this case $S_\alpha(x_0)$ has
a minimum at $\alpha \rightarrow 1$.  In oder to do this, we will present
solution (\ref{e9.b}) in a more convenient form, which allows to
separate leading singularities at $z\rightarrow 0$. After some
transformations involving hypergeometric functions \cite{stegun}, we
can rewrite Eq.~(\ref{e9.b}) as follows:
\begin{equation}
\label{e10.b}
S_\alpha(x_0)={2L(z-z^2)^{\alpha\over 2}\over \alpha -1}\left[
{M^{\alpha-1}\sin\left({\pi\alpha\over 2}\right)\over \pi}
-{(1-z)f_1(\alpha,z)\over \alpha \mbox{B}({\alpha\over 2},
{\alpha\over 2})}\right]
\end{equation} 
\[
-{L(2-\alpha)\over 2(\alpha-1)}\left[1-f_2(\alpha,z)(1-z)^{{\alpha \over 2}+1}
-{4 f_3(\alpha,z)z^{{\alpha\over 2}+1}
\over \alpha(\alpha+2)\mbox{B}({\alpha \over 2},{\alpha \over 2})}\right],
\]
where 
\[
f_1(\alpha,z)=F(\alpha,2,{\alpha\over 2}+1,z),
\]
\[
f_2(\alpha,z)=F(2-{\alpha\over 2},{\alpha\over 2},1-{\alpha\over 2},z),
\]
and
\[
f_3(\alpha,z)=F(2-{\alpha\over 2},{\alpha\over 2},{\alpha\over 2}+2,z).
\]
The first term in solution (\ref{e10.b}) decreases as $z^{\alpha / 2}$,
when $z \rightarrow 0$, while the second part decreases as $z$. 
If we take the starting point $x_0=r\ell_0$, where $r$ is constant, then 
$z=r/M$, where $M$ is large number, and we can separate the leading (with respect
to $M$) terms in solution  (\ref{e10.b}):
\begin{equation}
\label{e12.b}
S_\alpha(r\ell_0)=r^{\alpha / 2}{L\over \alpha -1}\left[
\eta(\alpha)M^{{\alpha\over 2}-1}
-\zeta(\alpha)M^{-{\alpha\over 2}}\right]
+L\cdot O(M^{-1}),
\end{equation} 
where 
\begin{equation}
\label{e13.b}
\eta(\alpha)={2\sin\left({\pi\alpha / 2}\right)\over \pi},~~~\quad
\zeta(\alpha)={2\over \alpha \mbox{B}({\alpha\over 2},{\alpha\over 2})}.
\end{equation} 
The above approximation accurately follows the solutions of the
discrete problem for $\alpha<1$, when the term
$\zeta(\alpha)M^{-{\alpha / 2}} $ dominates, but strongly deviate
from that for $\alpha >1$, when the term $\eta(\alpha)M^{{\alpha\over
2}-1}$ dominates (see Fig. 4).  The reason for these deviations is the
truncation of the non-leading terms in the Eq.~(\ref{e15a}).

In contrast with the discrete solution and
Monte-Carlo simulations of \cite{vis2}, expression Eq. (\ref{e12.b}) has
two minima: one at $\alpha=1+\varepsilon(M)$ and another at
$\alpha=2$. We will show that $\varepsilon(M)\rightarrow 0$ as
$M\rightarrow\infty$.  Let us expand $\eta$ and $\zeta$ in powers of
$\varepsilon$:
\begin{equation}
\label{e14.b}
\eta(1+\varepsilon)=\eta_0+\eta_1\varepsilon+ ...,~~~\quad
\zeta(1+\varepsilon)=\zeta_0+\zeta_1\varepsilon+ ... 
\end{equation}
Note that $\eta_0=\zeta_0=2/\pi$, and hence the expression
(\ref{e12.b}) does not have a singularity at $\alpha=1$. The location
of the minimum can be found by differentiation of the expansion for
$S_{1+\varepsilon}(r\ell_0)$ with respect to $\varepsilon$ and
equating the leading terms of the order of $\ln M$:
\begin{equation}
\label{e15.b}
\varepsilon=-{6(\eta_1+\zeta_1) +3\eta_0 \ln r \over \eta_0 (\ln M)^2}+ 
o\left( [\ln M]^{-2}\right)=
{6-12\ln 2 -3\ln r \over (\ln M)^2} +o\left( [\ln M]^{-2}\right).
\end{equation}
Indeed, Eq.~(\ref{e15.b}) shows that $\varepsilon(M)\rightarrow 0$ as
$M\rightarrow\infty$ and $r$ stays constant.

This analysis holds for any continuous functions $\eta$ and $\zeta$,
so long as $\eta(1)=\zeta(1)$ and, therefore, is likely to be valid in
the discrete case, in which, functions $\eta(\alpha)=\eta_d(\alpha)$
and $\zeta(\alpha)=\zeta_d(\alpha)$ do not satisfy
Eq.~(\ref{e13.b}). Note that $\eta_d(\alpha)$ can be expressed in
terms of function $\chi(\alpha)$ shown in Fig~3., namely
$\eta_d(\alpha)=\alpha\chi(\alpha)$. Analysis of Fig~3. shows that
$\chi(2) >0$.  Consequently, the minimum at $\alpha=2$ does not exist
in the discrete case. 

\section{Summary}

We have studied L\'evy Flights in a finite interval with absorbing
boundaries.  In Section II, we have derived expressions
Eqs.~(\ref{e6}) and (\ref{e2.4}) for the average total number of
flights (mean first passage time).  We also obtain a general recursion
relation Eq.~(\ref{e2.4a}) for the average of the sum of arbitrary
contributions from each flight in the form of the Fredholm integral
equation of the second kind.  We applied this method to derive the
probability of absorption by one of the boundaries.  In section III,
we have derived expressions Eqs.~(\ref{e2.5}) and (\ref{e2.3}) for the
average total path length of the L\'evy flights which is equivalent to
the mean first passage time of the L\'evy walks.

In Section IV, we have shown (See Appendix A) how the discrete L\'evy
flights are related to the fractional differential equation
Eq.~(\ref{e9}) of the superdiffusion with Riesz operator
Eq.~(\ref{e7c}).  For the continuous process described by
Eq.~(\ref{e9}), we derived exact analytical expressions
Eqs.~(\ref{e15b}), (\ref{e9.b}), and (\ref{e16.b}) for the mean first
passage time, the average total path length, and the probability of
absorption by one of the boundaries, respectively. All these
quantities are the solutions (See Appendix B) of the fractional
differential equation Eq. (\ref{e14a}) with Riesz kernel and with
different right hand sides.  In Section V, we have compared these
analytical solutions with numerical solutions obtained for the
discrete L\'evy flights (See Figs. 1-3). We have shown that fractional
differential formalism provides good approximation for the discrete
L\'evy flights in the interval with absorbing boundaries except the
case of $\alpha \rightarrow 2$ and the case when the starting point is
in the vicinity of an absorbing boundary. In the latter case the
fractional differential formalism yields correct scaling behavior with
respect to the interval size and distance to the boundary, but gives
an incorrect proportionality coefficient (See Fig. 3).

In Section VI, we have investigated the behavior of the
average path length as a function of the starting point and as a
function of $\alpha$. We have derived asymptotic expression Eq.
(\ref{e12.b}) for this quantity in the case when the starting point is located
close to the absorbing boundary.
We have shown that the expression for the average path length has a minimum
at $\alpha \approx 1$ if the process starts in the vicinity of the
absorbing boundaries (See Figs. 4,5).  This result, 
as well as Eqs. (\ref{e16.b}) and (\ref{e10.b}),
can be applied to the problem of light transmission through
cloudy atmosphere \cite{davis,pfe}.  

Similar fractional integral operators \cite{klafter,fog} --- namely the Riesz
operator $\nabla ^\alpha$ --- can be used to treat the problem of the
L\'evy flyer in the dimensions higher than one with randomly
distributed absorbing traps. Let $L$ be a characteristic distance
between neighboring traps. Then we still expect that the average
number of flights before absorption scales as $L^{\alpha}$ if the
process starts far away from the absorbing boundary and as
$L^{\alpha/2}$ if the process starts in the vicinity of the absorbing
boundary. This result is sufficient to prove that the minimum of the
average total path length traveled by the flyer before absorption is
achieved at $\alpha \rightarrow 1$, if the flyer starts in the
vicinity of the absorbing point.

Finally, we comment on the relevance of our findings to biological
L\'evy flight foraging. Our results essentially confirm that L\'evy
flights with $\alpha=1$ (or $\mu=2$ in notation of \cite{vis2}) should
theoretically provide the optimal strategy of foraging in case of
sparsely and randomly located food sites, if any food site can be
revisited many times \cite{vis2}. The presence of
the second minimum near $\alpha=2$ predicted by continuous limit
approximation may indicate another possible
strategy for foraging, i.e to perform Brownian walks in the region of
possible appearance of prey.  Break down of the 
continuous limit approximation in the vicinity of the absorbing boundary
indicates that the results should depend on the particular details
of the model.

\section{Acknowledgments}
We are grateful to M. F. Shlesinger, J. Klafter, C. Tsallis, A. Marshak,
A. Davis, and M. Gitterman for helpful suggestions and critical comments.
We also wish to thank an anonymous referee for constructive critique that
has helped to improve our manuscript.   
We thank NSF and CNPq for financial support.

\appendix

\section{Existence of the Continuous Limit Operator}

We will show that operator ${\cal D}_\alpha$ is defined on any function
$f(x)$ that has finite limits at both ends of the interval $f(0)$
and $f(L)$ and finite second derivative $f''(x)$ at any inner point $x$ 
of the interval $[0,L]$. According to
Eqs.~(\ref{e2}) and (\ref{e7}),
\begin{equation}
\label{e17.a}
{\cal D}_\alpha f(y)=\lim_{\ell_0\to 0}\ell_0^{-\alpha}\left\{{\alpha
\ell_0^\alpha\over 2}\left[\int_0^{y-\ell_0}{f(x)dx\over(y-x)^{\alpha+1}}+
\int_{y+\ell_0}^L {f(x)dx\over(x-y)^{\alpha+1}}\right]-f(y)\right\}.
\end{equation}
Making partial integration of both integrals in Eq.~(\ref{e17.a}) we get 
\[
{\cal D}_\alpha f(y) =
\lim_{\ell_0\to 0}\left\{\ell_0^{-\alpha}\left[{f(y-\ell_0)\over
2}+{f(y+\ell_0)\over 2}-f(y)\right] \right .
\]
\begin{equation}
\label{e18.a} 
\left .
+{1\over
2}\left[-\int_0^{y-\ell_0}{f'(x)dx\over(y-x)^{\alpha}}+
\int_{y+\ell_0}^L {f'(x)dx\over(x-y)^{\alpha}}\right]\right\}
\end{equation}
\[
-{f(0) \over 2y^\alpha} - {f(L) \over 2(L-y)^\alpha}.
\]
For $\alpha<2$, the first term in Eq.~(\ref{e18.a}) goes to zero as 
\[
{1\over 2}\ell_0^{2-\alpha}f''(y).
\]
The second term converges to the integral
\begin{equation}
\label{e19.a}
I \equiv \mbox{V.P.}\int_0^L{\mbox{sgn}(x-y)f'(x)dx\over 2|y-x|^\alpha},
\end{equation}
which exists for $\alpha<2$ if $f'(x)$ has
a derivative at $x=y$.  

Subtracting Eq.~(\ref{e19.a}) from the
second term in Eq.~(\ref{e18.a}), replacing $f'(x)$ in the integrand by
its Taylor expansion $f'(x)=f'(y)+f''(y)(x-y) +o(x-y)$, and combining it
with the first term of Eq.~(\ref{e18.a}), we reproduce Eq.~(\ref{e7b}) for
the correction operator $d_\alpha(\ell_0)$.  This shows that the
operator ${\cal D}_\alpha$ is well defined for the class of functions
with existing second derivative.

\section{Sonin Inversion Formula}

Equation (\ref{e14a}) belongs to a class of generalized Abel equations.
In his classical works, N. Ya. Sonin  \cite{sonin} 
suggested a general method for solving such equations. 
In particular \cite{popov}, an equation: 
\begin{equation}
\int_0^L\frac{\left[ a_1\mbox{sgn}\left( x-y\right) +a_2\right] ^\nu }
{2\mid x-y\mid^\alpha }\varphi (y)dy=h(x),~~~\quad 0\leq x\leq L  
\label{e1.b}
\end{equation}
has a solution: 
\begin{equation}
\varphi (z)=B_\alpha z ^{\beta -1}\frac d{dz }\int_z ^L\frac{%
t ^{1-\alpha }}{\left( t -z \right) ^{\beta -\alpha }dt}\frac d{dt}
\int_0^t \frac{y ^{\alpha -\beta }}{\left( t -y \right)
^{1-\beta }}h(y)dy ,  \label{e2.b}
\end{equation}
where
\begin{equation}
B_\alpha =-2\sin \left( \pi \beta \right) \Gamma \left( \alpha \right) \Gamma
^{-1}(\beta )\Gamma ^{-1}\left( 1-\beta +\alpha \right) \left( a_1+a_2\right)
^{-\nu }\pi ^{-1},  \label{e3.b}
\end{equation}
and parameter $\beta $ is determined 
by relations 
\begin{equation}
\sin \left( \pi (\beta -\alpha )\right) =c_\nu \sin \left( \pi \beta \right) ,
\label{e4.b}
\end{equation}

\begin{equation}
c_\nu =\left( \frac{a_2-a_1}{a_1+a_2}\right) ^\nu .  \label{e5.b}
\end{equation}
Similar inversion formulae can be found in ref. \cite{minsk}.

In case of Eq.~(\ref{e14a}), $\nu =1,a_2=0,a_1=-1.$ Hence, according to
Eq.~(\ref{e5.b}) $c_\nu =-1$.  Equation~(\ref{e4.b}) has infinite
number of solutions $\beta=\alpha/2 +k,$ where $k$ is an integer. For
$0<\alpha<2$, only two solutions with $k=0,k=1$ lead to the converging
integrals in Eq.~(\ref{e2.b}):

\begin{equation}
\label{e6.b}
\varphi_1(x)=B_\alpha z^{\alpha/2-1}{d\over dz}
\int_z^L dt~t^{1-\alpha}(t-z)^{\alpha/2}{d\over dt}
\int_0^t y^{\alpha/2}(t-y)^{\alpha/2-1}h(y)dy 
\end{equation}
and 
\begin{equation}
\label{e7.b}
\varphi_2(x)=-B_\alpha z^{\alpha/2}{d\over dz}
\int_z^L dt~t^{1-\alpha}(t-z)^{\alpha/2-1}{d\over dt}
\int_0^t y^{\alpha/2-1}(t-y)^{\alpha/2}h(y)dy, 
\end{equation}
where
\begin{equation} 
\label{e8.b}
B_\alpha={4\sin({\pi\alpha\over 2}) \over\pi\alpha \mbox{B}({\alpha
\over 2},{\alpha \over 2})}. 
\end{equation}
Since Eq.~(\ref{e14a}) contains
$f'(x)=\varphi(x)$, one can always satisfy the first boundary
condition Eq.~(\ref{e14b}) by defining $f(x)=\int_0 ^x \varphi(z)dz$.
Adding solution (\ref{e6.b}) for $h(x)=-1$ and solution (\ref{e7.b})
for $h(x)=1$, one can see that the homogeneous equation with $h(x)=0$
has a nontrivial solution $\varphi_0=(Lx-x^2)^{{\alpha\over
2}-1}$. 
Hence the second boundary condition Eq.~(\ref{e14b}) can be
satisfied if we select $\varphi=\varphi_1-C\varphi_0$, with constant
$C=L^{1-\alpha}\int_0^L\varphi(x)dx/\mbox{B}({\alpha \over 2},{\alpha
\over 2})$.  In case $h(x)=-1$, straightforward calculations lead to
Eq.~(\ref{e12}).

Now we will obtain the analytical solution for the average total path
length before absorption in the continuous process, $S_\alpha(x)$.  In this
case, the right hand side of Eq.~(\ref{e14a}), $h(x)=s(x)$, is given
by Eq.~(\ref{e2.2}), and we can use its symmetry $h(x)=h(L-x)$.  Thus,
in order to satisfy the second boundary condition Eq.~(\ref{e14b}), we
should have $\varphi(x)=-\varphi(L-x)$. To construct such a solution,
we first find the solution $\varphi_1(x)$ for the first term in
Eq.~(\ref{e2.2}) $h(x)=x^{1-\alpha}$.  Obviously, the function
$-\varphi_1(L-x)$ provides the solution for the second term
$h(x)=(L-x)^{1-\alpha}$. The solution for the third constant term is
given by Eq.~(\ref{e12}) with a proper coefficient.  Summing up all
three partial solutions and using various properties of hypergeometric
functions \cite{stegun}, one can find the total solution presented in
Eq.~(\ref{e9.b}).

\begin{figure}
\vspace{.2in}
\centerline{\psfig{figure=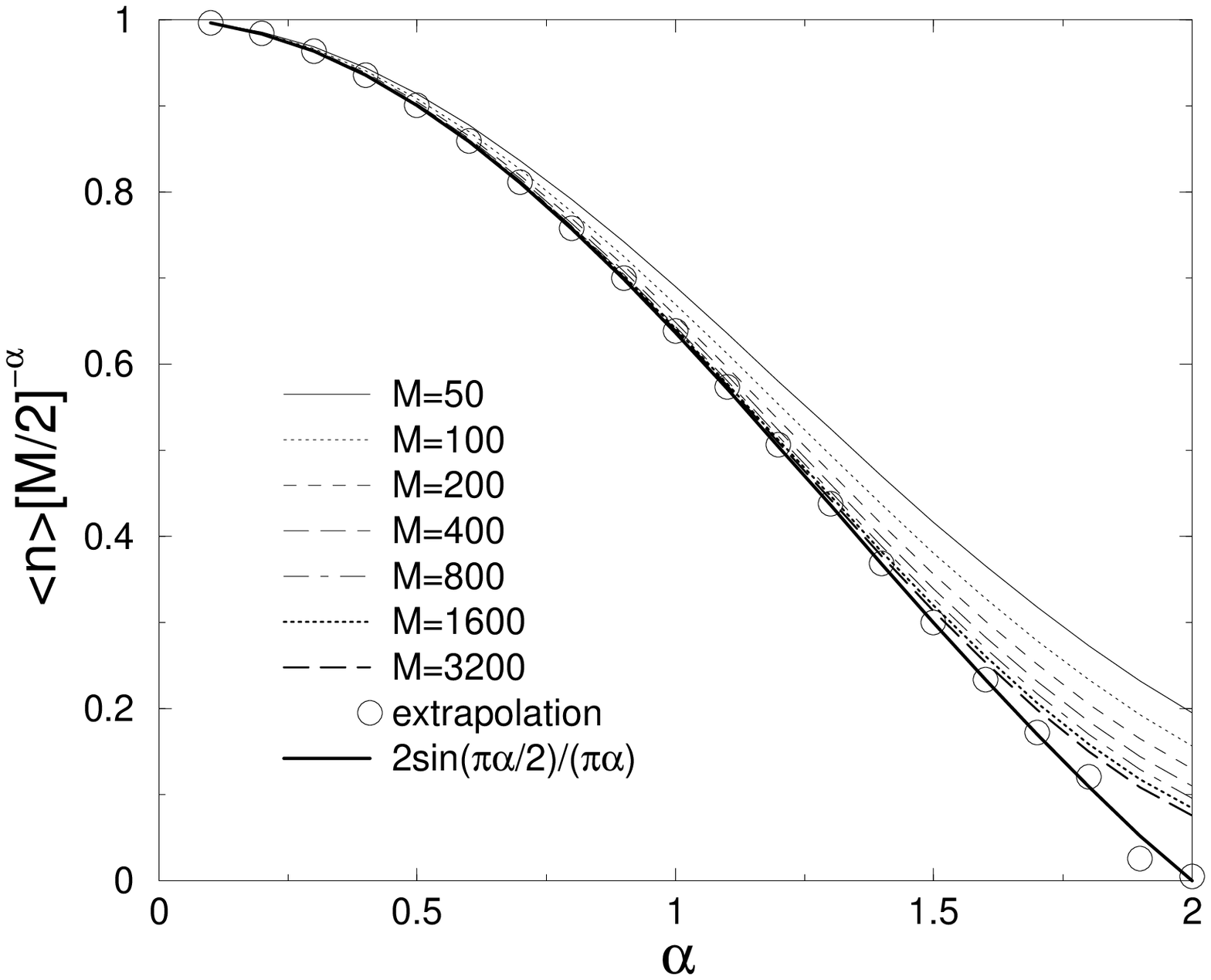,height=6truein}}
\caption{
The behavior of the scaled average number of flights $\langle
n\rangle (M/2)^{-\alpha}$ versus $\alpha$ for increasing values of
$M=50,100,200,400,800,1600,3200$ in the case $x_0=L/2$
in comparison with the continuous limit prediction of
Eq.~(\protect\ref{e15b}), $2\sin(\pi\alpha/2)/(\pi\alpha)$, shown as a
bold line.  We see good convergence to the predicted
function except for the values of $\alpha \approx 2$. We extrapolate
the values $\langle n\rangle (M/2)^{-\alpha}$ for $M \rightarrow \infty$
(circles) using their polynomial fits with respect to $M^{-1}$
for $\alpha \leq 1$, or with respect to $M^{2-\alpha}$ for $1<\alpha<2$,
or with respect to $(\ln M)^{-1}$ for $\alpha=2$.}
\label{fig3}
\end{figure}

\begin{figure}
\vspace{.2in}
\centerline{\psfig{figure=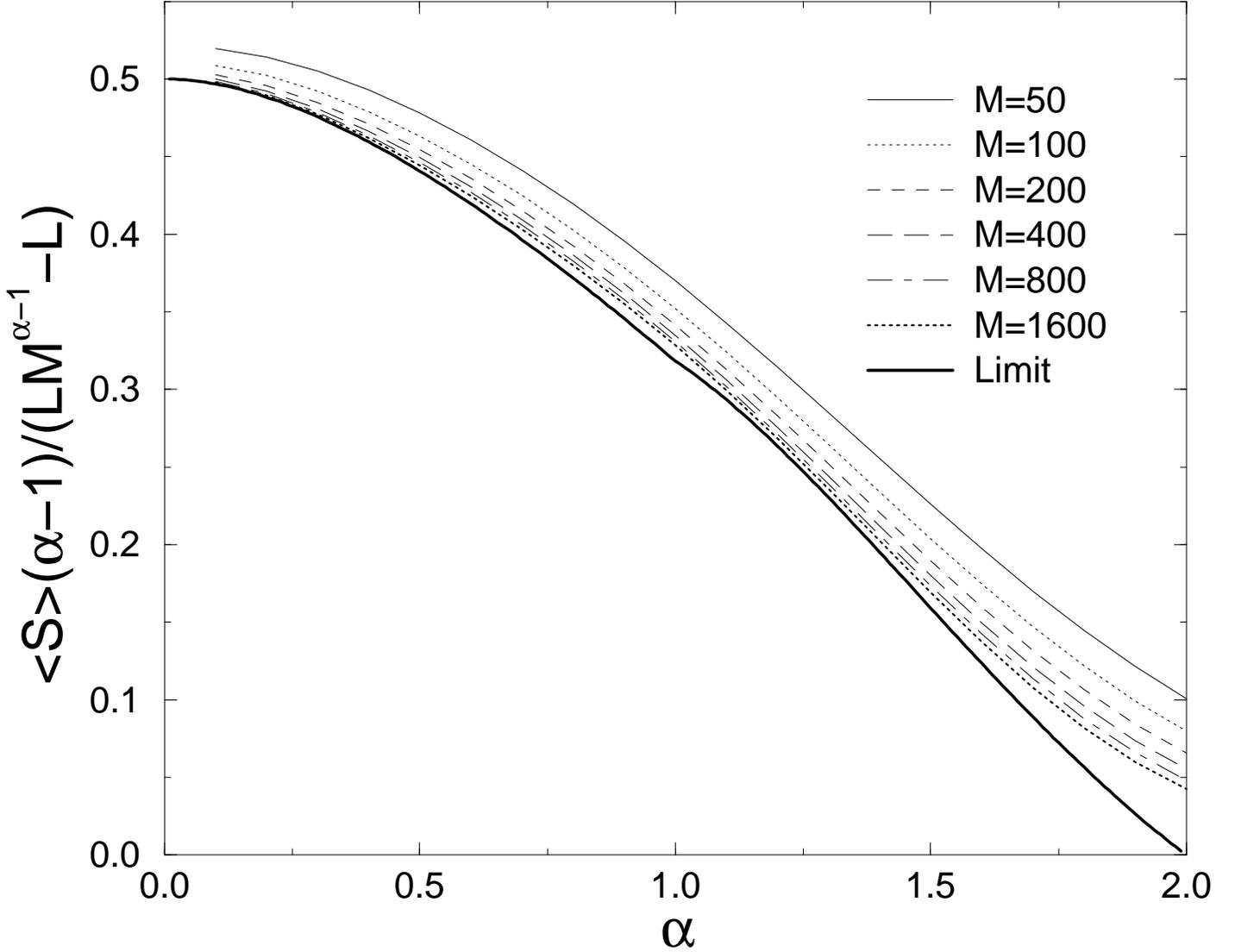,height=6truein}}
\caption{
The behavior of the scaled average total path length $\langle
S\rangle (\alpha-1)/(LM^{\alpha-1}-L)$ versus $\alpha$ for increasing
values of $M=50,100,200,400,800,1600$ in the case $x_0=L/2$ in
comparison with the continuous limit prediction of
Eq.~(\protect\ref{e9.b}) in the limit $M\rightarrow \infty$ shown as a
bold line. In this limit, continuous approximation follows first
term of Eq.~(\protect\ref{e9.b}) for $\alpha <1$ and the second term
$\sin(\pi\alpha/2)/\pi$ for $\alpha \geq 1$.  We see good convergence
to the predicted function except for the values of $\alpha \approx
2$.}
\label{fig5}
\end{figure}

\begin{figure}
\vspace{.2in}
\centerline{\psfig{figure=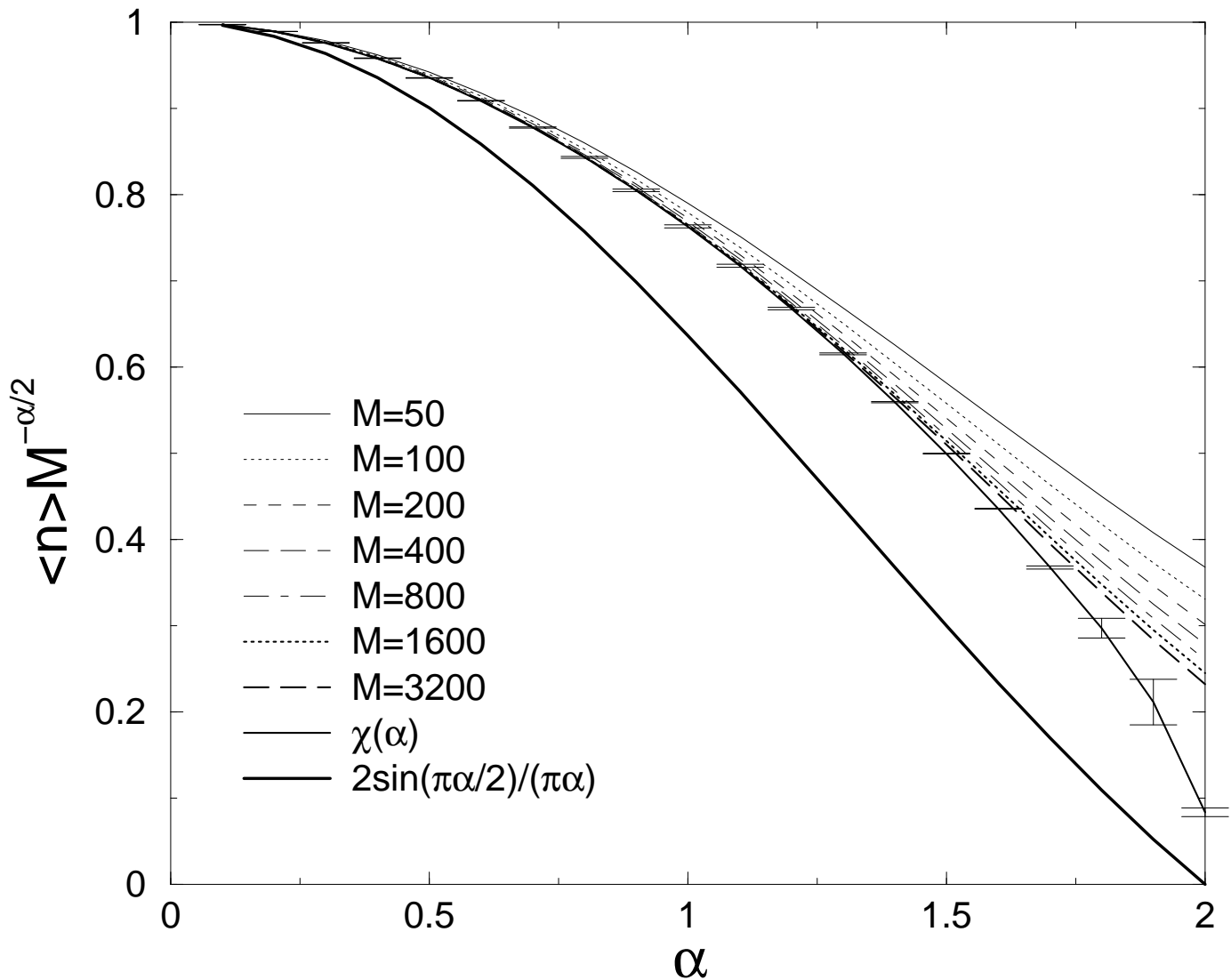,height=6truein}}
\caption{ The behavior of the scaled average number of flights
$\langle n\rangle (M)^{-\alpha/2}$ versus $\alpha$ for increasing
values of $M=50,100,200,400,800,1600,3200$ in the case $x_0=\ell_0$ in
comparison with the continuous limit prediction of
Eq.~(\protect\ref{e15b}) $2\sin(\pi\alpha/2)/(\pi\alpha)$, shown as a
bold line. Although the values are close to the continuous limit
predictions, they converge to a different function $\chi(\alpha)$ as
$M\rightarrow \infty$.  To obtain $\chi(\alpha)$, we extrapolate the
values $\langle n\rangle (M/2)^{-\alpha}$ for $M \rightarrow \infty$
using the same procedure as in Fig.~1. We assume that the error bars
are equal the discrepancies between the extrapolation
and the continuous limit in Fig.~1.}
\label{fig4}
\end{figure}

\begin{figure}
\vspace{.2in}
\centerline{\psfig{figure=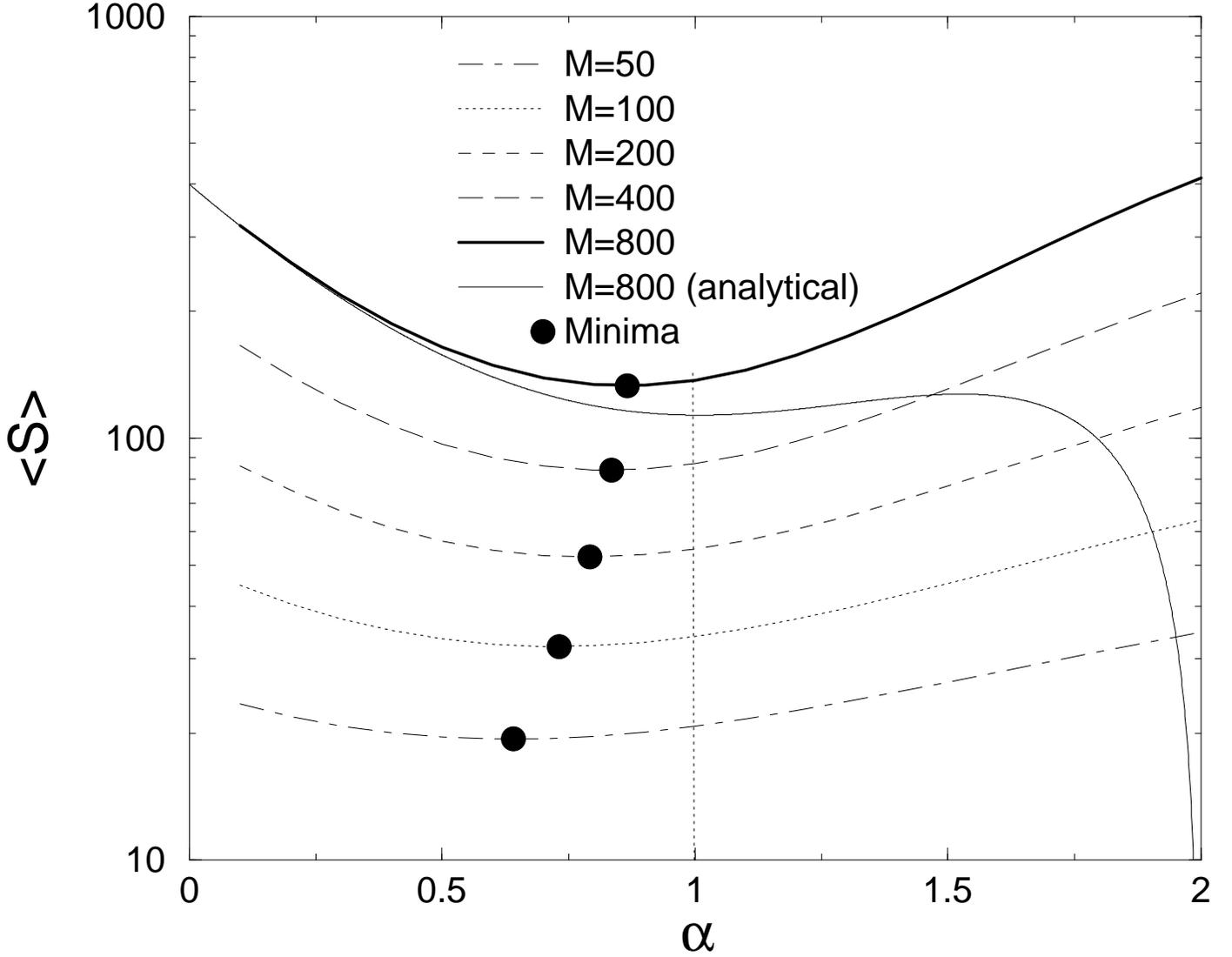,height=6truein}}
\caption{Semi-logarithmic plot of numerical solutions of 
$\langle S\rangle$ versus
$\alpha$ for the case $x_0=\ell_0=1$ and various values of
$M=50,100,200,400,800$. Circles indicate the
positions of the minima $\alpha_{\min}(M)$ which shift towards the
vertical line $\alpha=1$, as $M$ increases. In addition, we show the
analytical continuous limit approximation $S_\alpha(\ell_0)$ 
given by Eq.~(\protect\ref{e13.b}) 
for $M=800$.}
\label{fig1}
\end{figure}

\begin{figure}
\vspace{.2in}
\centerline{\psfig{figure=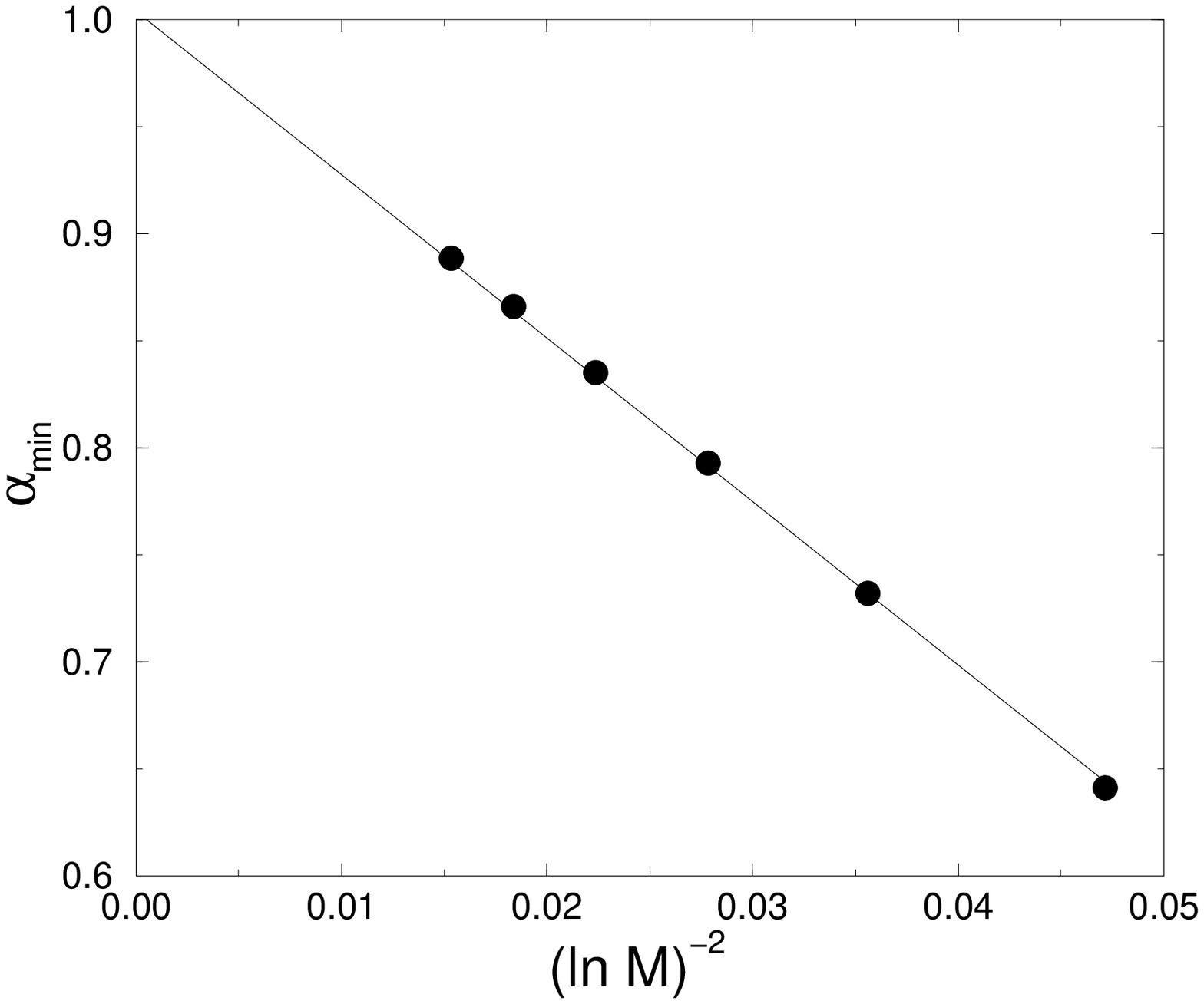,height=6truein}}
\caption{
The values of $\alpha_{\min}(M)$, determined in Fig.~4 as a 
function of $(\ln M)^{-2}$. The line shows linear least square fit,
obtained by including a hypothetical limiting value 
$\alpha_{\min}(\infty)=1$. }
\label{fig2}
\end{figure}


\begin{thebibliography}{MT1}

\bibitem{book1}
B. D. Hughes, {\it Random Walks and Random Environments, Volume 1:
Random Walks\/} (Clarendon Press, Oxford, 1995).
 
\bibitem{book2}
M. F. Shlesinger, G. Zaslavsky, and U. Frisch, eds., {\it L\'evy Flights
and Related Topics in Physics\/} (Springer, Berlin, 1995).

\bibitem{klafter}
R. Metzler and J. Klafter, Phys. Rep. {\bf 339}, 1 (2000). 

\bibitem{zolotarev}
V. M. Zolotarev and V. M. Uchaikin, {\it Chance and Stability: 
Stable Distributions and their Applications\/} (VSP BV, Utrecht, 1999). 


\bibitem{tsallis}
C. Tsallis, Phys. World {\bf 10}, 42 (1997)

\bibitem{west1}
B. J. West and V. Seshadri, Physica A {\bf 113}, 203 (1982).

\bibitem{shles}
M. F. Shlesinger and J. Klafter in 
{\it On Growth and Form}, edited by H. E. Stanley
and N. Ostrowsky (Nijhoff, Dordrecht, 1986), p.~283.

\bibitem{west2}
B. J. West and W. Deering, Phys. Reports {\bf 246}, 1 (1994).

\bibitem{zaslav2}
G. M. Zaslavsky, {\it Physics of Chaos in Hamiltonian System\/}
(Imperial College Press, London, 1998)

\bibitem{kutner}
R. Kutner, Physica A {\bf 264}, 84 (1999), R. Kutner, A. Pekalski,
K. Sznajd-Weron (Eds.), {\it Anomalous Diffusion: from Basics 
to Applications\/} (Springer-Verlag, Berlin 1999).

\bibitem{taqqu}
G. Samorodnitsky and M. S. Taqqu, {\it Stable Non-Gaussian Random Processes \/}
(Chapman \& Hall, New York, 1994) 

\bibitem{zaslav} 
G. M. Zaslavsky, Physica D {\bf 76}, 110 (1994).


\bibitem{fog}
H. C. Fogedby, Phys. Rev. E. {\bf 58}, 1690 (1998).

\bibitem{jesp}
S. Jespersen, R. Metzler, and H. C. Fogedby, Phys. Rev. E. {\bf 59},
2736 (1999).

\bibitem{saichev}
A. I. Saichev and G. M. Zaslavsky, Chaos {\bf 7}, 753 (1997).

\bibitem{gitter}
M. Gitterman, Phys. Rev. E {\bf 62}, 6065 (2000).


\bibitem{schneider}
W. R. Schneider and W. Wyss, J. Math, Phys. {\bf 30}, 134 (1989).
 
\bibitem{barkai}
E. Barkai and R.J. Sibley, J. Phys. Chem. B {\bf 104}, 3866 (2000).

\bibitem{mbk}
R. Metzler, E. Barkai, and J. Klafter, Phys. Rev. Lett. {\bf 82} 3563 (1999).

\bibitem{met1}
R. Metzler and J. Klafter, Physica A {\bf 278}, 107 (2000).

\bibitem{met2} 
R. Metzler and J. Klafter, Europhys. Lett. {\bf 51}, 492 (2000).
 
\bibitem{rang}
G. Rangarajan and M. Ding, Phys. Rev. E. {\bf 62}, 120 (2000). 

\bibitem{barkai1}
E. Barkai, Phys. Rev. E. {\bf 63}, 046118 (2001).

\bibitem{podlub}
I. Podlubny, {\it Fractional Differential Equations\/} (Academic Press,
London, 1999). 

\bibitem{rob}
P. M. Drysdale and P. A. Robinson, Phys. Rev. E {\bf 58}, 5382 (1998).

\bibitem{davis}
A. Davis and A. Marshak in {\it Fractal Frontiers \/} edited by 
M. M. Novak and T. G. Dewey (World Sci. River Edge, N.J.,1997) p.~63.

\bibitem{pfe}
K. Pfeilsticker, J. Geophys. Res. D {\bf 104} 4101 (1999).

\bibitem{smirnov}
V. I. Smirnov, 
{\it A course of higher mathematics, vol. 4:  
Integral equations \/} (Pergamon Press, Oxford, 1964).

\bibitem{minsk}
S. G. Samko, A. A. Kilbas, and O. I. Maritchev,
{\it Integrals and Derivatives of the Fractional Order and Some of 
Their Applications\/} (Nauka i Tekhnika, Minsk, 1987) [in Russian];
S. G. Samko, A. A. Kilbas, and O. I. Maritchev,
{\it  Fractional Integrals and Derivatives\/} (Gordon and Breach, Newark 1993).

\bibitem{arutun}
N. Kh. Arutyunyan, Prikl. Mat. Mekh. {\bf 23}, 901 (1959) [in Russian].

\bibitem{popov}
G. Ya. Popov, {\it Stress Construction near Punches, Cuts, Thin
Inclusions, and Supporters\/} (Nauka, Moskow, 1982) [in Russian].

\bibitem{sonin}
N. Ya. Sonin,
{\it Studies on cylinder functions and special polynomials\/}
(Gostekhizdat, Moscow, 1954) [in Russian].

\bibitem{levand}
M. Levandowsky, J. Klafter, and B. S. White, Bull. Mar. Sci. {\bf 43},
758 (1988).

\bibitem{cole}
B. J. Cole, Anim. Behav. {\bf 50}, 1317 (1995).

\bibitem{vis1}
G. M. Viswanathan, V. Afanasyev, S. V. Buldyrev, E. J. Murphy,
P. A. Prince and H.~E.~Stanley, Nature {\bf 381}, 413 (1996).

\bibitem{vis2} 
G. M. Viswanathan, S. Buldyrev, S. Havlin, M. G. E. da Luz, E. P. Raposo
and H. E. Stanley, Nature {\bf 401}, 911 (1999).

\bibitem{vis3}
G. M. Viswanathan, V. Afanasyev, S. V. Buldyrev, S. Havlin, M. G. E. da
Luz, E. P. Raposo, and H. E. Stanley, Physica A {\bf 282}, 1  (2000).


\bibitem{upad}
A. Upadhyaya, J.-P. Rieu, J. A. Glazier, and Y. Sawada,  
Physica A {\bf 293}, 549 (2001). 

\bibitem{stegun}
M. Abramowitz and I. A. Stegun, {\it Handbook of Mathematical Functions\/}
(Dover Publications Inc., New York, 1965); V. I. Smirnov, 
{\it A course of higher mathematics, vol. 3, part 2:  
Complex variables, special functions \/} (Pergamon Press, Oxford, 1964). 

\bibitem{kogan}
H. M. Kogan J. Uspekhi Mat. Nauk, {\bf 19} no 4, 228 (1964). [In Russian]

\bibitem{bg}
S. V. Buldyrev and M. Gitterman (unpublished).

\bibitem{D}
Note that one can define ${\cal D}_\alpha \equiv\lim_{\ell_0\to
0}\ell_0^{-2}[{\cal L}_\alpha(\ell_0)-{\cal I}]={\alpha\over
2(\alpha-2)}{d^2\over dx^2}$ for $\alpha>2$, and ${\cal D}_\alpha \equiv
\lim_{\ell_0\to 0}\ell_0^{-2}(-\ln \ell_0)^{-1} [{\cal
L}_\alpha(\ell_0)-{\cal I}]={d^2\over dx^2}$ for $\alpha=2$.

\end{thebibliography}
\end{document}